\documentclass[12pt]{JHEP3}
\usepackage{mathrsfs}
\usepackage{amsmath,amssymb}
\usepackage{relsize}
\usepackage{graphicx}

\def\({\left(}
\def\){\right)}

\def\sl(2){\alg{sl}(2)}

\def\be{\begin{equation}}
\def\ee{\end{equation}}

\newcommand{\bea}{\begin{eqnarray}}
\newcommand{\eea}{\end{eqnarray}}

\def\la{\label}

\def\ov{\over}

\newcommand{\alg}[1]{\mathfrak{#1}}

\newcommand{\AdS}{{\rm  AdS}_5\times {\rm S}^5}

\newcommand{\sfrac}[2]{{\textstyle\frac{#1}{#2}}}

\newcommand{\bem}{\left (\begin{matrix}}
\newcommand{\eem}{\end{matrix} \right )}

\def\hstar{\,\hat{\star}\,}

\author{Gleb Arutyunov$^a$\footnote{Email: G.E.Arutyunov@uu.nl, frolovs@maths.tcd.ie, rsuzuki@maths.tcd.ie} {}\footnote{Correspondent fellow at Steklov
Mathematical Institute, Moscow.}\,, \  Sergey Frolov$^{b\,\dagger}$
\, and\,  Ryo Suzuki$^b$
 \\ $^{a}$ {\it Institute for Theoretical
Physics and Spinoza Institute,\\ ~~Utrecht University, 3508 TD
Utrecht, The Netherlands} \\ $^b$ {\it Hamilton Mathematics Institute and School of Mathematics, \\
~~Trinity College, Dublin 2, Ireland} }

\abstract{ We use the Thermodynamic Bethe Ansatz equations for the
$\AdS$ mirror model to  derive the five-loop anomalous dimension
of the Konishi operator. We show numerically that the
corresponding result perfectly agrees with the one recently
obtained via the generalized L\"uscher formulae. This constitutes
an important test of the AdS/CFT TBA system.}

\title{Five-loop Konishi from the Mirror TBA}

\preprint{
          \smaller{\smaller{\smaller{ITP-UU-10-04}}}\\[-.5ex]
          \smaller{\smaller{\smaller{SPIN-10-04}}}\\[-.5ex]
          \smaller{\smaller{\smaller{TCDMATH 10-01}}}\\[-.5ex]
          \smaller{\smaller{\smaller{HMI-10-01}}}}
\begin{document}

\renewcommand{\thefootnote}{\arabic{footnote}}
\setcounter{footnote}{0}


\section{Introduction}

The mirror Thermodynamic Bethe Ansatz (TBA) is generalization of the relativistic TBA
\cite{Zamolodchikov90} and offers a tool to determine the spectrum
of the $\AdS$ superstring \cite{AJK,AF07}. Recently there has been
interesting progress on its precise formulation and in deriving some consequences of the corresponding spectral equations
\cite{AF09a}-\cite{Arutyunov:2009ax}. In parallel development the
four-loop anomalous dimension of the Konishi operator was obtained
\cite{BJ08} by means of the generalized L\"uscher formulae
\cite{Luscher85,JL07,BJ08}\footnote{See  \cite{HS08a}-\cite{BJ09}
for other interesting applications of L\"uscher's approach.} and
it exhibits a perfect match with a direct field-theoretic computation \cite{Sieg,Vel}.
More recently, a refined version of the generalized L\"uscher formula has been proposed, and applied to the five-loop anomalous dimension of the Konishi operator \cite{BJ09},
\begin{equation*}
\Delta^{(10)} = \Delta_{\rm asympt}^{(10)}
+ g^{10} \Big\{
-\frac{81 \zeta(3)^2}{16}+\frac{81 \zeta(3)}{32}-\frac{45 \zeta(5)}{4}+\frac{945 \zeta(7)}{32}-\frac{2835}{256} \Big\},
\end{equation*}
where $g$ is the coupling constant (string tension) related to
the 't Hooft coupling $\lambda$ through $\lambda=4\pi^2g^2$.
This result has been further generalized to all twist two operators
by also invoking the reciprocity principle \cite{Lukowski:2009ce}.
Although no field-theoretical computation has been done at five loops,
those anomalous dimension being continued to the negative values of
spin enjoy a
quite non-trivial agreement with the constraints imposed by the
BFKL equation \cite{Lipatov:1993yb}-\cite{KL02}.

\smallskip

Having in mind all these developments, it is time to ask whether
the proposed TBA spectral equations are in accord with the
perturbative findings based on the generalization of L\"uscher's
approach. As for Konishi at four loops, incorporation of the
corresponding result in the TBA framework does not pose any real
difficulty because the leading finite-size correction has been
built-in when the excited-state TBA is formulated; all the
contribution to the anomalous dimension (the energy of the
corresponding string state) comes from the main $Y_Q$-functions
taken at their asymptotic values \cite{GKV09}, the latter being
given by the generalized L\"uscher formulae constructed through
the infinite-volume scattering data \cite{BJ09}. Also, rapidities
$u_k$, $k=1,\ldots, N$, of the excited string particles forming an
$N$-particle state under consideration are determined by the
asymptotic Bethe Ansatz (ABA) equations (also known as the
Bethe-Yang equations).

\smallskip

The situations change, however, for the five-loop case. As was
argued in \cite{BJ08,BJ09}, to find the anomalous dimension of Konishi at five loops,
one has to compute the correction to the ABA, though still
the asymptotic $Y_Q$-functions can be used.
The exact Bethe equation determines the shift
$\delta u_k$ of particle rapidities $u_k=u_k^o+g^8 \delta u_k$ from their
asymptotic values $u_k^o$ as
$$
\sum_{j=1}^M \frac{\delta {\rm ABA}(u_k)}{\delta u_j}\, \delta u_j
+\Phi_k^{(8)}=0\, .
$$
The first term here is a variation of the ABA equations and the
term $\Phi^{(8)}$ is the leading finite-size correction to the ABA
of order $g^8$. The formula above should be evaluated at
$u_k=u_k^{(2)}$, where $u_k^{(2)}$ are the one-loop values of
the particle rapidities. As soon as $\delta u_k$ are determined,
the five-loop correction to the dimension (energy of the
corresponding string state) follows from expanding the exact
dispersion relation up to the five-loop order, {\it i.e.} up to
$g^{10}$.

\smallskip

It turns out that the correction $\Phi^{(8)}$ derived through analytic continuation of the mirror TBA equation looks rather different from the one derived
through L\"uscher-type perturbative arguments \cite{BJ09}. The
difference occurs due to the fact that the exact Bethe equations
involve auxiliary Y-functions which must satisfy a coupled system of the
TBA equations. Starting from five loops, the auxiliary Y-functions start to
contribute non-trivially to the modification of the ABA.

\smallskip

In this note we find a strong evidence that the results obtained
for the Konishi operator from both the TBA and L\"uscher
approaches are in an excellent agreement at the five-loop level.
Fortunately, as discussed above, what we have to do is to show
only that the correction to the ABA derived from the TBA agrees
with $\Phi^{(8)}$ found from L\"uscher's approach \cite{BJ09}.
This will be done by expanding the TBA equations around the
asymptotic solution \cite{GKV09} and by linearizing the exact
Bethe equations around the ABA. As we will see, the exact Bethe
equations at order $g^8$ involve a leading correction to a single
auxiliary function $Y_{1|vw}$, which we will determine from the
linearized TBA equations numerically. Then, using this result for
$Y_{1|vw}$ we evaluate numerically the correction to the exact
Bethe equations and find that it agrees with $\Phi^{(8)}$ of
\cite{BJ09} with a sufficiently high precision. It would be
important to support this numerical agreement by an analytic
proof.

\smallskip

The paper is organized as follows. In the next section we present
the linearization of the simplified TBA equations. We use them
in section 3 to compute numerically the correction to the exact Bethe
equations finding an agreement with the corresponding result in
\cite{BJ09}.

\section{Linearizing the TBA system}
In the light-cone gauge the string vacuum corresponds to the gauge
theory operator ${\rm tr} Z^J$, where $Z$ is one of the three
complex scalars of the ${\cal N}=4$ super Yang-Mills theory. In
this work we will be interested in the excited states from the
$\sl(2)$-sector, the latter comprises the composite operators of
the type  ${\rm tr}\, D^NZ^J$, where $D$ is a light-cone
derivative. For $J=2$ these are operators of twist two spin $N$.
In particular, the operator with $J=2$ and $N=2$ is the $\sl(2)$
descendent of the Konishi operator.

\smallskip

The Thermodynamic Bethe Ansatz approach for the $\AdS$ mirror
model leads to the following expression for the energy of the
corresponding $N$-particle string states \bea \label{fsE}
E=J+\sum_{i=1}^N{\cal
E}(p_i)-\frac{1}{2\pi}\sum_{Q=1}^{\infty}\int_{-\infty}^{\infty}{\rm
d }u\frac{d\widetilde{p}^Q}{du}\log(1+Y_Q)\, . \eea Here $J$ is
the angular momentum carried by the string rotating around the
equator of ${\rm S}^5$. The integration runs over a real rapidity
line of the mirror theory, $\widetilde{p}^Q$ and $Y_Q$ are momenta
and Y-functions of the mirror $Q$-particles. The asymptotic
energies of string theory particles with momenta $p_i$ are fixed
by the dispersion relation \cite{Beisert:2004hm, Beisert05}

\bea {\cal E}(p)=\sqrt{1+4g^2\sin^2\frac{p}{2}}\, . \eea The
function $Y_Q$ is exponentially small at large $J$.  Therefore the
last term of \eqref{fsE} can be regarded as a finite-size
correction to the asymptotic spectrum of string energies.

\smallskip

For a fixed $J$ and small $g$, the $Y_Q$ functions become small again.
The finite-size corrections provide so-called wrapping corrections to the energy
or the anomalous dimension at weak coupling.
In particular, for the Konishi operator the finite-size
effects make their appearance starting from $g^8$ that corresponds
to the fourth loop order of perturbation theory \cite{BJ08}.

\smallskip

Recall that the large $J$ asymptotic solution of the excited-state TBA equations can correctly reproduce the leading finite-size corrections to the energy. The TBA equations are formulated in terms of the following Y-functions: $Y_Q$-functions
associated with $Q$-particle bound states, auxiliary functions
$Y_{Q|vw}$ for $Q|vw$-strings, $Y_{Q|w}$ for $Q|w$-strings, and
$Y_{\pm}$ for $y_\pm$-particles \cite{AF09a,AF09b}.

\smallskip

In what follows it is convenient to use the simplified TBA
equations derived in \cite{AF09b,AF09d,Arutyunov:2009ax}. To
determine the leading finite-size correction to the asymptotic
form of $Y$-functions, we linearize the simplified TBA equations by
introducing for any Y-function the following representation
\bea
Y=Y^{o}(1+{\mathscr Y}) \,, \eea where $Y^{o}$ is the corresponding
asymptotic expression and ${\mathscr Y}$ is treated as the
perturbation. The linearized TBA equations will then take the
following form

\bigskip
 \noindent
$\bullet$\ $M|w$-strings: $\ M\ge 1\ $, ${\mathscr Y}_{0|w}=0$
\bea\la{Yforws} {\mathscr Y}_{M|w}=  (A_{M-1|w}{\mathscr
Y}_{M-1|w}+ A_{M+1|w}{\mathscr Y}_{M+1|w})\star s +\delta_{M1}\,
\Big(\frac{{\mathscr Y}_+}{1-Y_+^o}-\frac{{\mathscr
Y}_-}{1-Y_-^o}\Big)\hstar s\, ,~~~~~ \eea where we have introduced
the concise notation $A_{M|w}=\frac{Y^o_{M|w}}{1+Y^o_{M|w}}$.

\bigskip
 \noindent
$\bullet$\ $M|vw$-strings: $\ M\ge 1\ $, ${\mathscr Y}_{0|vw}=0$
\bea\la{Yforvw3} &&{\mathscr Y}_{M|vw}=  (A_{M-1|vw}{\mathscr
Y}_{M-1|vw}+ A_{M+1|vw}{\mathscr Y}_{M+1|vw})\star s
-Y_{M+1}^o\star s \\
\nonumber &&\hspace{8cm}+\, \delta_{M1}\, \Big(\frac{{\mathscr
Y}_-}{1-\frac{1}{Y_-^o}}-\frac{{\mathscr Y}_+}{1-\frac{1}{Y_+^o}}
\Big)\hstar s\, \,.~~~~~ \eea
Here we also defined $A_{M|vw}\equiv {Y_{M|vw}^o\ov 1+Y_{M|vw}^o }$.
Any asymptotic Y-functions except for $Y_Q^\circ$ have the magnitude of order 1, so $A_{M|w}\,, A_{M|vw}$ are not small.
Note that the Y-function $Y^o_{M}$ provides the leading large $J$
correction to the asymptotic TBA equations and, for this reason,
it enters the last equation as an inhomogeneous term.

\bigskip
 \noindent
$\bullet$\   $y$-particles \bea\la{Yfory1} &&{\mathscr Y}_+ -
{\mathscr Y}_-= Y_{Q}^o\star K_{Qy}\,,~~~~~~~ \\
\la{Yfory2}
 &&{\mathscr Y}_+ +
{\mathscr Y}_-= 2(A_{1|vw}{\mathscr Y}_{1|vw}-A_{1|w}{\mathscr
Y}_{1|w})\star s  -Y_Q^o\star s+2 Y_Q^o\star K_{xv}^{Q1}\star s\,
. \eea

We will not need the equations for $Q$-particles, because the asymptotic
solution for $Y_Q$ is already known from the generalized L\"uscher's formulas.
The equations for $Q$-particles suggest that the next corrections to $Y_{Q}^o$ start from $g^{16}$.

\smallskip

The linearized TBA equations above define the leading, exponential
in $J$ correction to the asymptotic Y-functions. However, by further fixing
$J=2$ we would like to view ${\mathscr Y}$ as
another source of perturbative correction in $g$ to the asymptotic Y-functions.
Therefore it makes sense to expand further the asymptotic Y-functions in
the linearized TBA equations in
powers of $g$ having in mind that expansion of $Y_M$ starts from
$g^8$.

The rapidities $u_k$ of the Konishi operator are expanded at small $g$ as
$u_1 = - u_2 = \frac{1}{\sqrt{3} \, g} \left[ 1 + 2 \, g^2 - \frac{5 \, g^4}{4} + {\cal O} (g^6) \right]$ using the notation of \cite{Arutyunov:2009ax}.
It is convenient to rescale them
as $u_k \to u_k/g $ and do the same $u\to u/g$
with the argument $u$ of the $Y_{M}^o(u)$, $Y_{M|vw}^o(u)$ and
$Y_{M|w}^o(u)$. On the other hand, the functions $Y_{\pm}^o(u)$ are
supported on the segment $u\in [-2,2]$ and as for them, the variable
$u$ will be kept unrescaled.

\smallskip

First, let us consider the equation for $y$-particles. Performing
the change of variables $u\to u/g$ in the
convolution term $Y_Q^o\star s$ and expanding in $g$, we see that
\bea
Y_Q^o \star s = \int_{-\infty}^{\infty} \, \frac{{\rm d}u}{g} \,
Y_Q^o \Big( \frac{u}{g} \Big)
\, \frac{g}{4\cosh\frac{\pi g}{2}(\frac{u}{g}-\frac{v}{g})}
= \int_{-\infty}^{\infty}{\rm d}u \, \frac{Y_Q^o(u)}{4\cosh\frac{\pi }{2}(u-v)} \,,
\eea
where the rescaled Y-function $Y_Q^o(u)$ reads as
{\small\bea \label{YQg8}
&& Y_Q^o(u)=g^8 \,
\frac{64 \, Q^2[-1+Q^2+u^2-w^2]^2}{(Q^2+u^2)^4[(Q-1)^2+(u-w)^2][(Q+1)^2+(u-w)^2]}\times
\\ \nonumber
&& \hspace{5cm}\times
\frac{1}{[(Q-1)^2+(u+w)^2][(Q+1)^2+(u+w)^2]} \ + {\cal O}
(g^{10}), \eea} \hspace{-0.255cm} \noindent and we denote by
$w\equiv u_1=-u_2 = \frac{1}{\sqrt{3}} + {\cal O} (g^2)$
the rescaled rapidity of the two-particle state
corresponding to the Konishi operator. Thus, $Y_Q\star s$ starts
at order $g^8$ in the weak-coupling expansion. Analogously, one
can establish that in eq.(\ref{Yfory1}) at $g\to 0$ one has
$Y_{Q}^o\star K_{Qy}\sim g^9$, while in eq.(\ref{Yfory2}) one
finds $Y_Q^o\star K_{xv}^{Q1}\star s\sim g^8$.  As a consequence,
eqs.(\ref{Yfory1}) and (\ref{Yfory2}) imply that the first
non-trivial corrections ${\mathscr Y}_{\pm}$ start at order $g^8$,
although for the difference one has ${\mathscr Y}_{+}-{\mathscr
Y}_{-}\sim g^9$.

\smallskip

In fact, there is a general pattern behind
inherited from the asymptotic solution for $Y_{\pm}^o$ ---
the sum ${\mathscr Y}_{+}+{\mathscr Y}_{-}$ admits an expansion
in even powers of $g$, while the difference
${\mathscr Y}_{+}-{\mathscr Y}_{-}$ in odd. Another interesting property
is that ${\mathscr Y}_{+}(g)={\mathscr Y}_{-}(-g)$. Thus, in the perturbative
expansion, the coefficients of ${\mathscr Y}_{\pm}$ in front of
even powers of $g$ coincide.

\smallskip

Now we turn our attention to eq.(\ref{Yforvw3}). The notation
$\hat{\star}$ signifies that integration in the corresponding
convolution term is over the segment $[-2,2]$.
Due to the fact that we do not
rescale the integration variable in the integrals
involving ${\mathscr Y}_{\pm}$, the expansion of $s$ in this
convolution term starts from $g$. Moreover, one finds that at the
leading order in $g$:
\bea
Y_+^o=Y_-^o=\frac{w^2-3}{w^2+1} \,,
\eea
which is a $u$-independent quantity. Hence,
\bea \nonumber &&
\Big(\frac{{\mathscr Y}_-}{1-\frac{1}{Y_-^o}}-\frac{{\mathscr
Y}_+}{1-\frac{1}{Y_+^o}} \Big)\hstar s\,
\stackrel{g\to 0}{=}
\frac{1}{4}(1+w^2)({\mathscr Y}_- - {\mathscr Y}_+)\hstar s
\sim g^{10}\,,
\eea
and, therefore this term produces higher than
the leading $g^8$ contribution into the equation for ${\mathscr
Y}_{M|vw}$.

\smallskip

Analogous consideration can be applied to eq.(\ref{Yforws}) which
shows that the correction ${\mathscr Y}_{M|w}$ starts from the
order $g^{10}$. Thus, at order $g^8$ we have the following
equations to be considered

\bigskip
 \noindent
$\bullet$\ $M|vw$-strings: $\ M\ge 1\ $, ${\mathscr Y}_{0|vw}=0$
\bea\la{Yforvw8l} &&{\mathscr Y}_{M|vw}=  (A_{M-1|vw}{\mathscr
Y}_{M-1|vw}+ A_{M+1|vw}{\mathscr Y}_{M+1|vw})\star s
-Y_{M+1}^o\star s .
\eea

\bigskip
 \noindent
$\bullet$\   $y$-particles ${\mathscr Y}_+={\mathscr Y}_-$
\bea \la{Yfory2s8l}
2{\mathscr Y}_+ = 2 A_{1|vw}{\mathscr
Y}_{1|vw}\star s -Y_Q^o\star s+2 Y_Q^o\star K_{xv}^{Q1}\star s\, ,
\eea where all the TBA kernels are taken at their leading order.
In fact, as we will see below, the correction to the asymptotic
Bethe-Yang equations involves at order $g^8$ the functions
${\mathscr Y}_{M|vw}$ alone, and therefore,
the only equation we have to solve is (\ref{Yforvw8l}).

\begin{figure}
\begin{center}
\includegraphics*[width=0.55\textwidth]{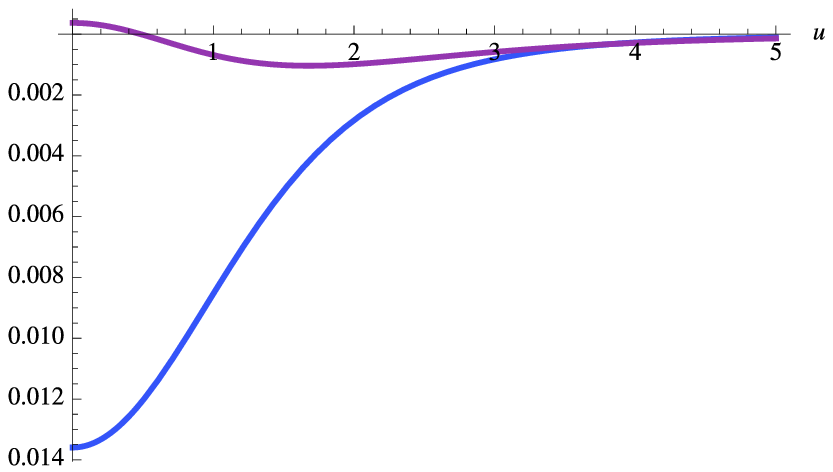}\\[2mm]
\hspace{-0.40cm}\includegraphics*[width=0.55\textwidth]{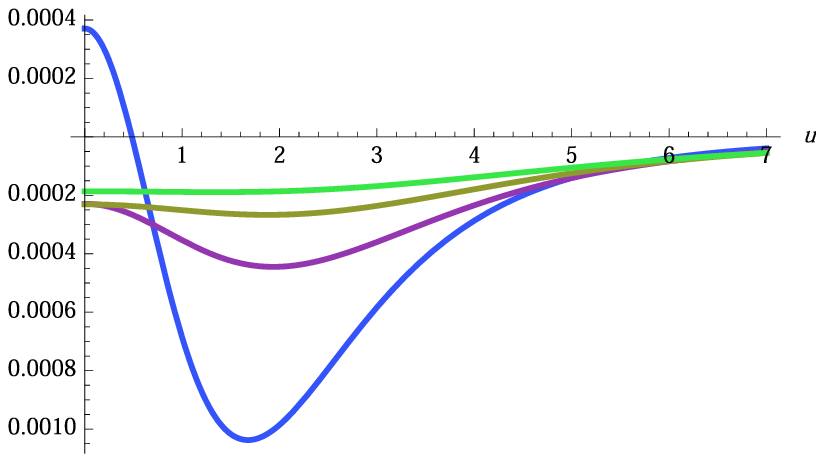}
\end{center}
\caption{In the upper picture the profiles of the functions
${\mathscr Y}_{1|vw}/g^8$ (blue) and ${\mathscr Y}_{2|vw}/g^8$ (purple) are depicted.
They are obtained by solving numerically eq.(2.13). The lower picture
contains the profiles of
${\mathscr Y}_{M|vw}/g^8$ for $M=2,3,4,5$. The absolute value of ${\mathscr Y}_{M|vw}$ decreases as $M$ increases. }
\end{figure}

\noindent Eq.(\ref{Yforvw8l}) can be equivalently written as
 \bea
 \label{ME}
{\mathscr Y}_{Q|vw}\star \Omega_{QM}= - Y_{M+1}^o\star s\, . \eea
Here we defined a kernel
\begin{align}
\Omega_{QM} (u,v) &= \delta_{QM} \, \delta(u-v)
\\[1mm]
&\quad - \delta_{Q,M-1} A_{M-1|vw} (u) \, s(u-v) - \delta_{Q,M+1} A_{M+1|vw} (u) \, s (u-v) \,,
\notag
\end{align}
where
\begin{small}
\bea && A_{M|vw}(u)=\frac{M(M+2)}{(M+1)^2}\times
\label{AMvw expand} \\
&& \hspace{2cm}\times
\Big(1+\frac{1+w^2}{\sqrt{M(M+2)-w^2}}\frac{1}{u^2+\rho_+^2}-
\frac{1+w^2}{\sqrt{M(M+2)-w^2}}\frac{1}{u^2+\rho_-^2}\Big)
+ {\cal O} (g^2),
\nonumber
\eea
\end{small} \hspace{-2.5mm}with $\rho_{\pm}^2=(1\pm
\sqrt{M(M+2)-w^2})^2$ and $s(u)=\frac{1}{4\cosh\frac{\pi u}{2}}$.
The formal solution to eq.(\ref{ME}) can be given with the help of
the inverse kernel $\Omega^{-1}$
\bea \label{YQvw}
{\mathscr Y}_{Q|vw}=- Y_{M+1}^o\star s\star \Omega^{-1}_{MQ}\, . \eea
 The kernel $\Omega$ can be inverted by
the power series expansion and computed numerically. This gives
rise to a numerical determination of the correction ${\mathscr
Y}_{1|vw}$. Alternatively, eq.(\ref{ME}) can be solved by
iterations, and it is what we have used.

We studied the following inhomogeneous linear equation
\begin{equation}
{\mathscr Y}_{M|vw}^{[n]} =  (A_{M-1|vw}{\mathscr
Y}_{M-1|vw}^{[n]} + A_{M+1|vw} {\mathscr Y}_{M+1|vw}^{[n]})\star s
-\delta_{n,M+1} Y_{M+1}^o\star s ,
\label{Split equation}
\end{equation}
which can be solved by numerical iterations starting  with
${\mathscr Y}_{M|vw}^{[n]}=0$ for each $n=2,3, \ldots$. Since all
equations are linear, the solution of the original equation
\eqref{ME} is given by the sum ${\mathscr Y}_{M|vw} \equiv
\sum_{n=2}^\infty {\mathscr Y}_{M|vw}^{[n]}$. The solution is
unambiguously well-defined provided that the homogeneous equation
has only the trivial solution ${\mathscr Y}_{M|vw}=0$. The
profiles of the first few ${\mathscr Y}_{M|vw}$ are presented in
Figure 1.

\section{Correction to the asymptotic Bethe equations}
The exact Bethe equations determine the exact (non-asymptotic)
positions of the Bethe roots $u_k$. As was shown in
\cite{Arutyunov:2009ax},  for Konishi-like states and below the
first critical point the exact Bethe equations\footnote{ According
to \cite{DT96, DT97}, the exact Bethe equations are
$Y_{1*}(u_k)=-1$, where $Y_{1*}$ is the function $Y_1$
analytically continued to the real rapidity line of the string
region.  The exact Bethe equations which follow from the canonical
TBA equation for $Y_1$ have been discussed in \cite{GKV09b}. We
exploit here the exact Bethe equations in a different
representation which is derived from the hybrid TBA equation for
$Y_1$ discussed in \cite{Arutyunov:2009ax}. } admit the following
representation \bea \pi (2n_k+1) &=&L\, p_k +i\sum_{j=1}^N \log
S_{\sl(2)}^{1_*1_*}(u_{j},u_{k}) +{\rm Im}~U \la{ExactBAEIm}\\
 &-& \nonumber 2\sum_{M=1}^{\infty}\log\left(1+Y_M
\right)\star \left( {\rm Im}\, K^{\Sigma}_{M1_*}
- s\star {\rm Im}\,K_{vwx}^{M-1,1_*} \right)  ~~~~~~
\\[1mm]
\nonumber
&+&2 \log(1+Y_{1|vw})\star
\left({\rm Im}\, \tilde{s}-i s\,\hat{\star}\,K^{y1_*}\right)
\nonumber  \\[1mm]
\nonumber &-&2\log{1-Y_-\ov 1-Y_+}\hstar s\star {\rm Im }\, K^{11_*}_{vwx}
-i\log\Big(1- \frac{1}{Y_-}\Big)\Big(1- \frac{1}{Y_+}\Big) \hstar K^{y1_*}
\,.
\eea
The kernels
$K^{\Sigma}_{M1_*}$, $K^{y1_*}$ and $K^{M1_*}_{vwx}$ with one leg
on the real rapidity line of the mirror theory and the other one
on the real line of string theory are given in
\cite{Arutyunov:2009ax}. The quantity $U$ appearing in the first
line of the exact Bethe equation is defined as
\bea  U(v) &\equiv
& 2\int_{-\infty}^{+\infty}{\rm d}t\,\log\Big[
\prod_{j=1}^N S(u_j^- -t)(t-u_j)\Big] K_{vwx}^{11*}(t,v)\\
\nonumber
&&\hspace{4cm}-2\sum_{j=1}^N\log(u_j-v-\frac{2i}{g})\frac{x_j^--\frac{1}{x^-}}{x_j^--\frac{1}{x^+}}\,
. \eea \normalsize\noindent
Finally, the kernel $\tilde{s}$ is defined as
\bea \tilde{s}(u,v) \equiv s(u-v-{i\ov g}+i0)\,.\eea
The second variable of any kernel entering eq.(\ref{ExactBAEIm})
and the function $U(v)$ should be evaluated at $v=u_k$.
Only the imaginary part of the exact Bethe equations is written down in eq.\eqref{ExactBAEIm}, because using the explicit form of the kernels one can recognize that the real part vanishes, as it should be for the equation to have a real solution.
Note that eq.(\ref{ExactBAEIm}) does not involve
$Y_{M|w}$-functions at all, and out of $Y_{M|vw}$ only $Y_{1|vw}$ is present.

\smallskip

The exact Bethe equations can be written as
 \bea \pi (2n_k+1)&=&J\,
p_k +i\sum_{j=1}^N \log S_{\sl(2)}^{1_*1_*}(u_{j},u_{k}) +{\cal
R}_k\, . \eea Now we would like to apply the same strategy as the one in
the TBA equations and find the leading weak coupling correction to
the asymptotic value ${\cal R}_k^{\rm asympt}=0$. The condition
${\cal R}_k^{\rm asympt}=0$ must be satisfied for any $g$ to
guarantee the compatibility of the exact Bethe equations with the
asymptotic Bethe ansatz. Taking into account that for the Konishi
state $J=2$ and $L=4$, we have checked numerically that this is
indeed the case,
for arbitrary values of $w \equiv u_1=-u_2$ inside the region where only $Y_{1|vw}$ has two zeroes in the strip $|{\rm Im} \, v|<1/g$.
Thanks to this identity, we can remove the term $\delta w \, \partial_w U$
from the corrections to the asymptotic Bethe ansatz.

Next, one can see that the leading weak-coupling
correction, which we denote by ${\cal R}_k$ starts at $g^8$.
It turns out that the last two lines in eq.(\ref{ExactBAEIm})
contain the convolution $\hat{\star}$ which, just as in the case
of the TBA equations, will not contribute to the leading order of
perturbation theory. Thus, at leading order we have the following
shift of the asymptotic Bethe-Yang equations
\begin{small}
\bea\label{cR} \delta{\cal R}_k=- 2\sum_{M=1}^{\infty}Y_M^o\star
{\rm Im}\,K^{\Sigma}_{M1_*} +2 A_{1|vw}{\mathscr Y}_{1|vw}\star
{\rm Im}\, \tilde{s} +2\sum_{M=1}^{\infty}Y_{M+1}^o\star s\star
{\rm Im}\, K_{vwx}^{11_*}\, .
\eea
\end{small} \hspace{-0.252cm} Here
all the kernels are evaluated at $g=0$ which provides the
leading order contribution. In particular, we have
\bea
{\rm Im}\, K^{11_*}_{vwx}=-\frac{1}{2\pi }\Big[\frac{u-v}{4+(u-v)^2}-{\rm
p.v.}\frac{1}{u-v} \Big]\, .
\eea
The convolution of the kernels in
the last term of eq.(\ref{cR}) can be computed in terms of Digamma
functions; for all $M\geq 1$ we find
\bea\nonumber s\star {\rm
Im}\, K^{M1_*}_{vwx}&=& \frac{1}{8\pi i
}\Big[-i\frac{4(u-v)}{M^2+(u-v)^2}\\ \nonumber
&&\hspace{1cm}+\,\psi\big(\sfrac{M}{4}+\sfrac{i}{4}(u-v)\big)-\psi\big(\sfrac{M}{4}-\sfrac{i}{4}(u-v)\big)\\
&&\hspace{1cm}-\,\psi\big(\sfrac{M+2}{4}+\sfrac{i}{4}(u-v)\big)+\psi\big(\sfrac{M+2}{4}-\sfrac{i}{4}(u-v)\big)\Big]\,
. \eea
Further, one needs ${\rm Im} \, \tilde{s}={\rm
p.v.}\frac{1}{4\sinh\frac{\pi}{2}(u-v)}$. Finally, the leading
contribution of the imaginary part of the dressing kernel in the
mirror-string region is given by
\bea {\rm Im}\,K^{\Sigma}_{M1_*}=
-\frac{u-v}{2\pi[(M+1)^2+(u-v)^2]}\, .
\eea
The $\Phi$ and $\Psi$ functions which appear in the definition of the dressing phase kernel \cite{AF09c}, do not contribute at this order, because they can be written as integrals over the interval $u \in [-2g,2g]$ after the rescaling of rapidities.
Combining now everything together, we read off the necessary
correction\footnote{In the second line of the equation below the
function $1/\sinh(u-u_k)$ exhibits singularity at $u=u_k$. We can
however omit the principal value prescription for the integral
because $A_{1|vw}$ vanishes at $u=u_k$ making the integrand
regular everywhere. } \bea \delta {\cal R}_k&=&
\sum_{M=1}^{\infty}\int_{-\infty}^{\infty}{\rm
d}u\, Y_M^o(u)\,\frac{1}{\pi}\frac{u-u_k}{(M+1)^2+(u-u_k)^2}\nonumber \\
\label{TBAr} &&\hspace{0.38cm} +\int_{-\infty}^{\infty} {\rm d}u\,
A_{1|vw}(u){\mathscr Y}_{1|vw}(u)\,\frac{1}{2\sinh\frac{\pi}{2}(u-u_k)}
\\  \nonumber
&+&\sum_{M=1}^{\infty}\int_{-\infty}^{\infty}{\rm
d}u\,Y_{M+1}^o(u)\,  \frac{1}{4\pi i
}\Big[-i\frac{4(u-u_k)}{M^2+(u-u_k)^2}\\ \nonumber
&&\hspace{4.7cm}+\,\psi\big(\sfrac{M}{4}+\sfrac{i}{4}(u-u_k)\big)-\psi\big(\sfrac{M}{4}-\sfrac{i}{4}(u-u_k)\big)\\
\nonumber
&&\hspace{4.7cm}-\,\psi\big(\sfrac{M+2}{4}+\sfrac{i}{4}(u-u_k)\big)+\psi\big(\sfrac{M+2}{4}-\sfrac{i}{4}(u-u_k)\big)\Big]\,
. \eea

In \cite{BJ09} the leading correction to the asymptotic Bethe-Yang
equations was denoted by $\Phi^{(8)}$. Comparison of our
definitions with that of \cite{BJ09} shows that the agreement
between the L\"uscher and the TBA approaches relies on the
fulfillment of the following equality \bea \delta {\cal
R}_k+\Phi^{(8)}(u_k)=0 \, ,\, \eea where
$u_1=w=\frac{1}{\sqrt{3}}=-u_2$ is the one-loop rapidity for the
two-particle state corresponding to the Konishi operator.

 According to \cite{BJ09}, the
quantity $\Phi^{(8)}$ was found to be \bea\label{Luscher}
&&\Phi^{(8)}(u_k)=\sum_{M=1}^{\infty}\int_{-\infty}^{\infty}{\rm d}u\,Y_M^{o}(u)\times\\
\nonumber &&~~~~~~\times \frac{1}{\pi}\Big[
-\frac{u-u_k}{(M+1)^2+(u-u_k)^2}-\frac{u-u_k}{(M-1)^2+(u-u_k)^2}+\frac{u_k}{-1+M^2+u^2-u_k^2}\Big]\,
. \eea Apparently, formulae eqs.(\ref{TBAr}) and (\ref{Luscher})
look rather different, only the first line in eq.(\ref{TBAr}),
which is due to the contribution of the dressing kernel, cancels
in the sum with the first term in eq.(\ref{Luscher}).

\smallskip

Plugging in eq.(\ref{TBAr}) the representation (\ref{YQvw})
specified for ${\mathscr Y}_{1|vw}$, we see that the last two
lines in (\ref{TBAr}) never involve $Y_1^o$. The last two terms in
eq.(\ref{Luscher}) exhibit the same phenomenon: for $M=1$ they
cancel each other. This suggests to compare individual terms in
eqs.(\ref{TBAr},\ref{Luscher}) associated with a contribution of a
given $Y_M$.
The case of $M=1$ trivially matches, and the term with ${\mathscr Y}_{1|vw}$ should be understood as ${\mathscr Y}_{1|vw}^{[M]}$ defined in \eqref{Split equation}.
Performing numerical evaluation of the corresponding
quantities, we summarized the results in Table 1 for $M=2,\ldots,
6$.

\bigskip

\begin{table}[htbp]
\begin{center}
Table 1. Comparison of the numerical results \vskip 0.3cm

 {\renewcommand{\arraystretch}{1.7}
\renewcommand{\tabcolsep}{0.22cm}
\begin{tabular}{|c||l|l|l|l|}
\hline $M$  & TBA & L\"uscher & TBA-L\"uscher & $\frac{\mbox{TBA-L\"uscher}}{\mbox{L\"uscher}}$ \\
\hline\hline
2   & $-0.0108303$ &  $-0.0108304$ & $5.21829\cdot 10^{-8}$ & $4.81819\cdot 10^{-6}$ \\
3   & $-0.000118621$ & $-0.000118665$ & $4.34822\cdot 10^{-8}$ &
0.000366429
\\
4   & $-4.63638\cdot 10^{-6}$ & $-4.6417\cdot 10^{-6}$ & $
5.32116\cdot 10^{-9}$ &
$0.00114638$  \\
5 & $-3.78671\cdot 10^{-7}$ & $ -3.79407\cdot 10^{-7}$ &
$ 7.35899\cdot 10^{-10}$  & $0.0019396$ \\
6 & $-4.89345\cdot 10^{-8}$ & $-4.94077\cdot 10^{-8} $ &
$4.73166\cdot 10^{-10}$ & $0.00957676$ \\
 \hline
\end{tabular}}
\vskip 0.5cm

\parbox{14.0cm}{\small The table describes a numerical comparison between $\delta {\cal R}_1$ and $-\Phi^{(8)}(u_1 )$.
In the column ``TBA" the results for the contributions to $\delta
{\cal R}_1$  produced by individual $Y_M$'s  are presented for
$M=2,\ldots, 6$. Analogously, the column ``L\"uscher" represents
the same quantities but for $-\Phi^{(8)}(u_1 )$. The contributions of the first line in (\ref{TBAr}) and the first term in (\ref{Luscher}) are not included in the table. The other two
columns give the absolute and relative estimates for the
difference between the TBA and L\"uscher approaches. }

\end{center}
\end{table}

\vskip 0.2cm

Albeit looking different, the individual $Y_M$-contributions show
a good agreement between $\delta {\cal R}_k$ and $-\Phi^{(8)}$.
Increasing $M$ the precision tends to decrease, but this is because
the contributions corresponding to higher $M$ become rather small
and their evaluation requires more time.

\smallskip

We conclude this paper with an observation to be understood in
future. On the one hand, the expression \eqref{Luscher} can be
concisely written as
\begin{equation}
\Phi^{(8)}(u_1) = - \Phi^{(8)}(u_2) = - \sum_{M=1}^{\infty}\int_{-\infty}^{\infty}
\frac{{\rm d}u}{4\pi} \, \partial_w Y_M^{o}(u) \,,
\end{equation}
by using \eqref{YQg8}. Here the level-matching condition is imposed before the differentiation.
The right hand side involves the derivative of the transfer matrix, and looks
similar to the known result for the ${\rm O}(4)$ model
\cite{GKV08}. On the other hand, the generalized L\"uscher formula
proposed in \cite{BJ09} involves the trace of the derivative of
the S-matrix with respect to the mirror rapidity $u$. The two quantities coincide for the present problem
of Konishi at five loops, but must be distinguished in the general
situation.\footnote{R.S. thanks Zoltan Bajnok and Nikolay Gromov
for a discussion on this point.}

\vskip 1cm

\section*{Acknowledgements}
We thank Zoltan Bajnok and Juan Maldacena for interesting
discussions. The work of G.~A. was supported in part by the RFBR
grant 08-01-00798-a, by the grant NSh-672.2006.1, by NWO grant
047017015 and by the INTAS contract 03-51-6346. The work of S.F.
was supported in part by the Science Foundation Ireland under
Grants No. 07/RFP/PHYF104 and 09/RFP/PHY2142. The work of R.S. was
supported  by the Science Foundation Ireland under Grants No.
07/RFP/PHYF104.


\end{document}